# Quantitative and Qualitative Evaluation of Explainable Deep Learning Methods for Ophthalmic Diagnosis

Authors:
Amitojdeep Singh[*,1,2], J. Jothi Balaji[3], Varadharajan Jayakumar[1], Mohammed Abdul Rasheed[1], Rajiv Raman[4], Vasudevan Lakshminarayanan[1,2]

[1]Theoretical and Experimental Epistemology Lab (TEEL), School of Optometry
[2]Department of Systems Design Engineering, University of Waterloo, Canada
[3]Department of Optometry, Medical Research Foundation, Chennai, India
[4]Shri Bhagwan Mahavir Vitreoretinal Services, Sankara Nethralaya, Chennai, India

* Amitojdeep Singh Brar, School of Optometry and Vision Science, University of Waterloo, 200 University Ave. W. Waterloo, ON, Canada, N2L 3G1. E-mail: amitojdeep.singh@uwaterloo.ca

## SYNOPSIS

Deep learning methods used for retinal diagnosis are typically "black boxes" that cannot explain how the system made its decision. In this study multiple explainability methods that highlight anomalous regions in OCT scans are compared for their clinical significance.

## ABSTRACT

**Background:** The lack of explanations for the decisions made by algorithms such as deep learning has hampered their acceptance by the clinical community despite highly accurate results on multiple problems. Recently, attribution methods have emerged for explaining deep learning models, and they have been tested on medical imaging problems. The performance of attribution methods is compared on standard machine learning datasets and not on medical images. In this study, we perform a comparative analysis to determine the most suitable explainability method for retinal OCT diagnosis.

**Methods:** A commonly used deep learning model known as Inception – v3 was trained to diagnose 3 retinal diseases - choroidal neovascularization (CNV), diabetic macular edema (DME), and drusen. The explanations from 13 different attribution methods were rated by a panel of 14 clinicians for clinical significance. Feedback was obtained from the clinicians regarding the current and future scope of such methods.

**Results:** An attribution method based on Taylor series expansion, called Deep Taylor was rated the highest by clinicians with a median rating of 3.85/5. It was followed by Guided backpropagation, and SHAP (SHapley Additive exPlanations).

**Conclusion:** Explanations of deep learning models can make them more transparent for clinical diagnosis. This study found that the best performing method for a medical diagnosis task such as retinal OCT diagnosis may not be the one considered best for other deep learning tasks. Overall, there was a high degree of acceptance from the clinicians surveyed in the study.

# INTRODUCTION

Retinal diseases are prevalent among large sections of the society, especially amongst the aging population and also those with other systemic diseases such as diabetes. In the United States, it is estimated that, 40 million people suffer from eye diseases which could lead to blindness if left untreated [1]. It is estimated that number of Americans over 40 years with a diabetic retinopathy (DR) diagnosis will rise three folds from 5.5 million in 2005 to 16 million in 2050 [2]. For each decade of age after 40, the prevalence of low vision and blindness increases by a factor of three [3]. Long wait times in the developed world and lack of access to healthcare in the developing countries lead to delays in diagnosis and in turn deteriorated vision and even blindness. This leads to financial burden (and psychological burden) on the patients as well as the healthcare system due to higher treatment costs in the later stages.

Artificial intelligence (AI), especially deep learning which is modelled after the human neural system [4] has produced promising results in many areas including ophthalmology. These are used for tasks like disease detection [5], segmentation [6], and quality enhancement of optical coherence tomography (OCT) and fundus photographs [7]. The convolutional neural networks (CNN) are the most common form of deep learning algorithms used for image classification tasks like disease detection. These have shown promising results for diagnosis of DR, AMD and glaucoma [1,8,9].

Even though these algorithms show performance comparable to that of human experts, the applications of DL methods in ophthalmology is limited. A major hindrance is the 'black-box' nature of these algorithms as they cannot explain how the algorithm arrived at that decision unlike a clinician. Additionally, these also face medico-legal and technical challenges which could be resolved with newer legislations, user-centric systems, and improved training.

Various explainability methods have been developed and applied to different areas including medical imaging [10]. Most of the studies, especially the ones for ophthalmic diagnosis utilize a single explainability method and do not provide comparisons with alternatives [11,12]. We argue that an explainability method that performs the best on standard computer vision datasets may not be the most suitable for OCT images which have a different data distribution than real-world images. In previous studies [13,14], we compared multiple explainability methods quantitatively for their ability to highlight the part of the image which had the most impact on the model decision. We did an exploratory qualitative analysis using ratings from 3 optometrists and the results showed the need for a more detailed analysis to judge these methods [14].

In this study, we compare and evaluate 13 explainable deep learning methods for diagnosis of three retinal conditions – choroidal neovascularization (CNV), diabetic macular edema (DME), and drusen. These methods were rated by a panel of 14 eye care professionals. Their observations regarding the clinical significance of these methods, preference regarding AI systems, and suggestions for future implementations are also analyzed herein.

# METHODS

In this section, we discuss the deep learning model used to detect the diseases along with a brief overview of the explainability methods used to generate the heatmaps of the regions the model considered for making the decisions.

## *Model*

A CNN called Inception-v3 [15] is used for many computer vision tasks including the diagnosis of retinal images was used to classify the data from the UCSD dataset [16] into 4 classes – CNV, DME, drusen, and normal. The model was trained on 84,000 images and tested on 1,000 images, 250 from each class resulting in a test accuracy of 99.3%. The confusion matrix showing the relationship between true and predicted classes is shown in Table 1. It compares the predicted label (diagnosis) by the model on the X-axis with true labels (ground truths) on the Y-axis.

## *Explainability with attributions*

Explaining a deep learning method is the exercise of determining the effect of each input feature, or each pixel of the image on the output of the model. These methods attribute the contribution of each pixel of the image to the model output and are hence called attribution methods. The results are provided as the contribution of each pixel and the visualizations are called heatmaps as they are brighter for pixels with a higher impact on the output.

The attribution methods used in this study can be categorized into 3 types apart from the baseline occlusion which involves covering parts of the image to see the impact on the output. The function-based methods derive attributions directly from the model gradients and include gradient and Smoothgrad [17]. The signal-based methods analyze the flow of information (signal) through layers of neural network and include DeConvNet [18], Guided BackPropagation (GBP) [19], and Saliency [20]. The methods based on attributions completely include Deep Taylor [21], DeepLIFT [22], Integrated Gradients (IG) [23], input times gradient, Layerwise Relevance Propagation [24] with Epsilon rule (LRP EPS), Layerwise Relevance Propagation with Z rule (LRP Z), and SHAP [25]. SHAP and Deep LIFT are considered as state-of-the-art on standard machine learning datasets and have superior theoretical background while IG is commonly used for retinal images in literature [11,12].

The heatmaps for 3 correctly and 1 incorrectly classified example of using the attribution methods are described in Figures 1 and 2. It must be noted that certain methods such as DeepTaylor and Saliency provide only positive evidence. Those providing both positive and negative evidence have some high-frequency noise of negative evidence that can be removed in practice but retained here to compare original outputs.

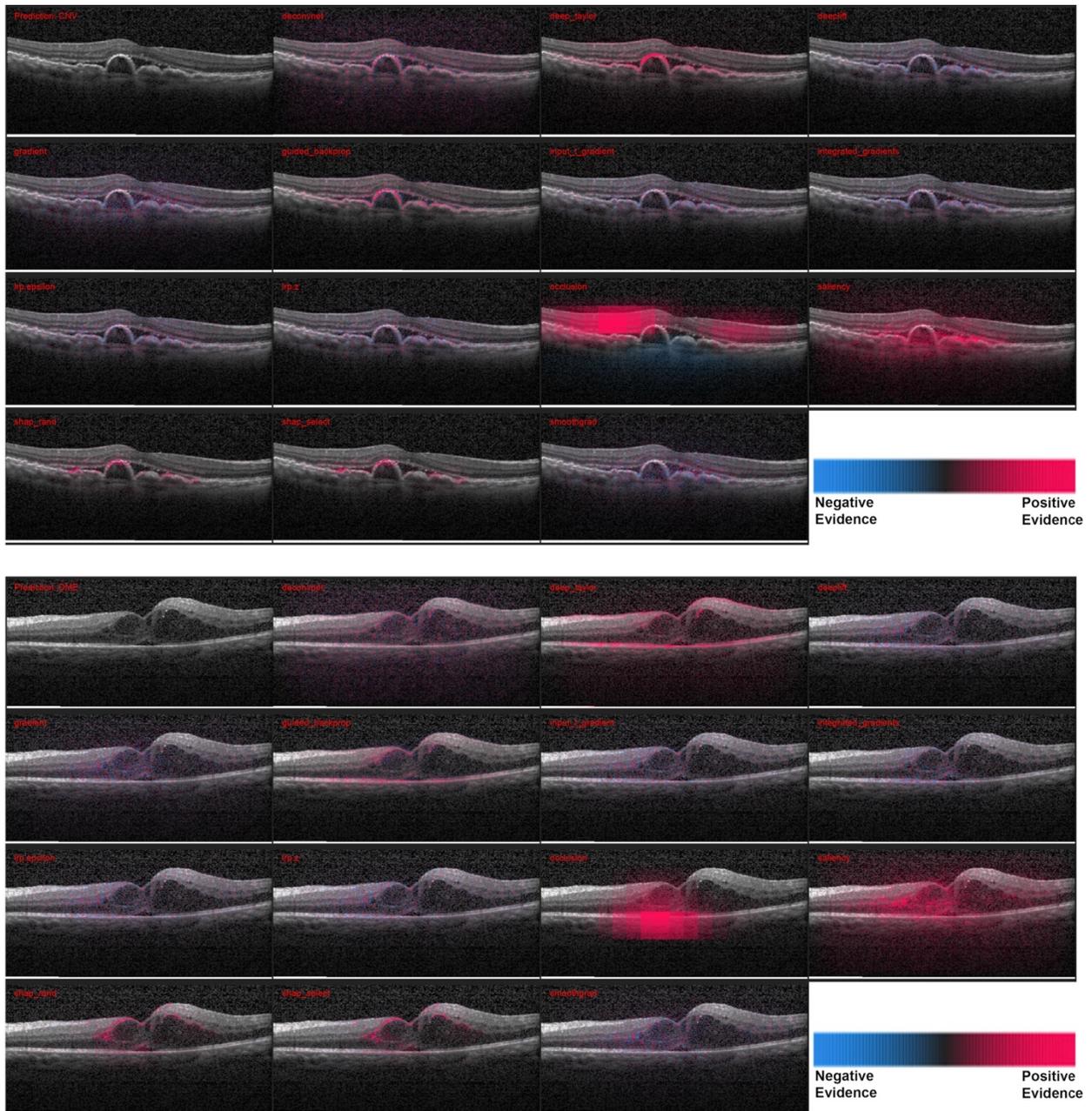

Figure 1: Heatmaps for scans with the larger pathologies - (top) choroidal neovascularization (CNV) and (bottom) diabetic macular edema (DME). For each case - Row 1: Input image, DeConvNet, Deep Taylor, DeepLIFT. Row 2: Gradient, GBP, Input times gradient, IG. Row 3: LRP – EPS, LRP – Z, Occlusion, Salience. Row 1: Input image, DeConvNet, Deep Taylor, DeepLIFT. Row 2: Gradient, GBP, Input times gradient, IG. Row 3: LRP – EPS, LRP – Z, Occlusion, Salience. Row 4: SHAP Random, SHAP Selected, SmoothGrad. The scale in the bottom right shows that the parts highlighted in magenta color provide positive evidence regarding presence of a disease while those in blue color provide a negative evidence indicating that the image is closer to normal. DeepTaylor, GBP perform the best, SHAP highlights partial but precise regions.

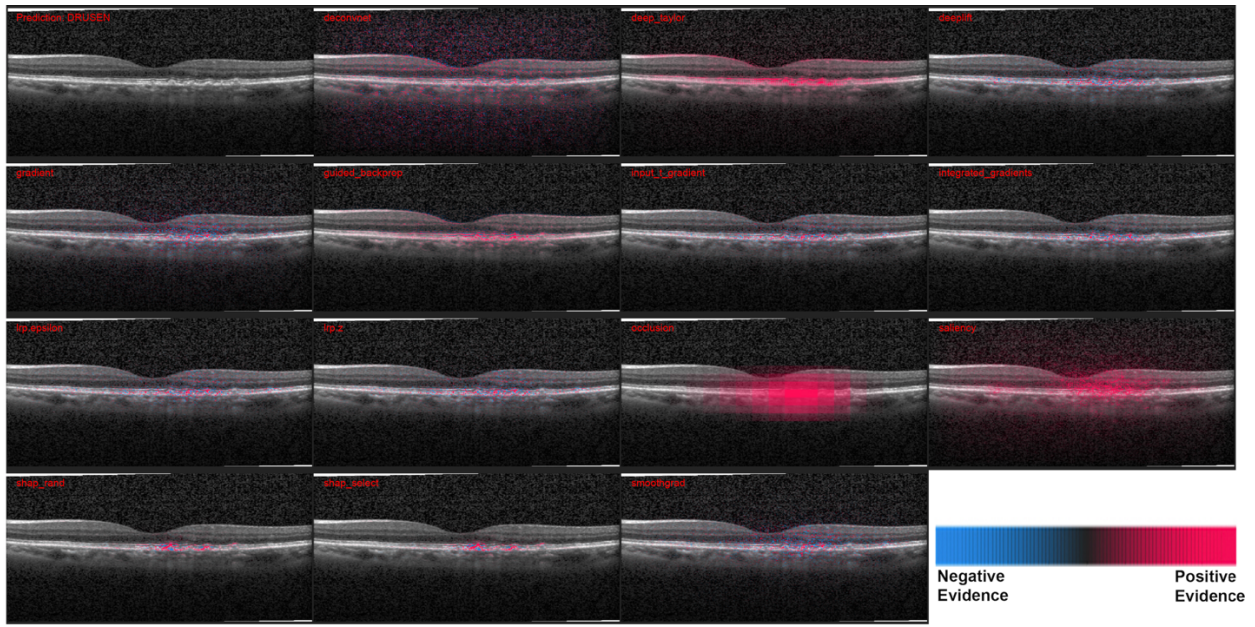

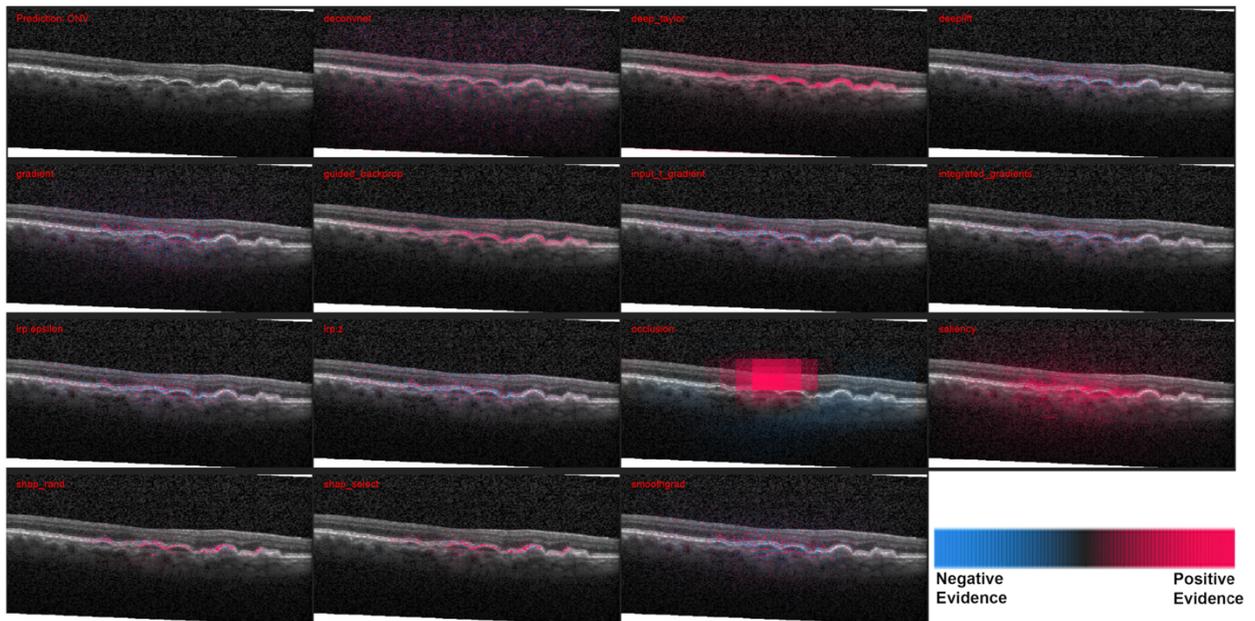

*Figure 2: Heatmaps for 2 scans with drusen, the smaller pathology. Top: Correct diagnosis, Bottom: Incorrect diagnosis. The pathological structures are smaller than the previous two and as a result most of the methods highlight regions outside too. SHAP is the most precise here in. In the incorrect case there is higher negative evidence (blue), especially with occlusion.*

*Table 1: Confusion matrix for the model on the test set of 1000 images*

|  |  | Predicted Label | | | | |
|---|---|---|---|---|---|---|
|  |  | CNV | DME | Drusen | Normal | Total |
| True Label | CNV | 249 | 0 | 1 | 0 | 250 |
|  | DME | 1 | 249 | 0 | 0 | 250 |
|  | Drusen | 3 | 0 | 247 | 0 | 250 |
|  | Normal | 0 | 0 | 2 | 248 | 250 |

# RESULTS

The heatmaps generated by the 13 methods for 20 images for each disease category were presented to a team of 14 clinicians – 10 ophthalmologists and 4 optometrists. The group had a median experience of 5 years with retinal diagnosis, out of which 4 years with OCT images. The average number of images rated per week was 40 with all the clinicians having prior experience analyzing retinal Spectral domain optical coherence tomography (SD-OCT) images. They rated the explanations from 0 to 5 with 0 being no relevance and 5 being fully relevant. The scores of each clinician were normalized by subtracting the respective mean and then rescaling between 0 to 5.

*Comparison between methods*

The violin plots of normalized scores of raters for all the methods across 60 scans are shown in Figure 3. The estimated probability density of each method is shown by the thickness of the violin plot. Table 2 gives the rating data for all conditions and methods. Deep Taylor with the highest median rating of 3.85 was judged as the best performing method. It is relatively simple to compute and involves Taylor series expansion of the signal at neurons. It was considerably ahead of GBP, the next best method which was closely followed by SHAP with selected and then random background.

IG, which is commonly employed in the literature for generating heatmaps for retinal diagnosis [11,12] received a median score of only 2.5. It is known to be strongly related and in some cases mathematically equivalent to LRP EPS [26] and this was also reflected in similar ratings. The Z rule of EPS was not found to make much difference and the simple to compute input times gradient performed reasonably well. DeepLIFT could not be tested in its newer Reveal Cancel rule due to compatibility issues with the model architecture and the older Rescale rule had a below average performance. As expected, the baseline occlusion which used sliding window of size 64 to cover the pixel and then compute significance performed worse than the attribution-based methods.

Most of the methods have the majority of the values around the median indicating consistent ratings across images and raters. Both cases of SHAP and Saliency have particularly elongated distributions. For SHAP, the curve is widest around 4 indicating good ratings for many cases.

However, the values around 2.5 due to lower coverage of pathology drive the overall median lower. In the case of Saliency, the ratings are spread from about 4.5 to 1.5 with many of them around 3.25 and 1.75 marks. The former is due to larger coverage of the pathological region and the latter is due to the fact that it missed regions frequently. Hence, despite better median value, it is not as suitable as lower-rated methods such as IG where a bulk of the value is around the median.

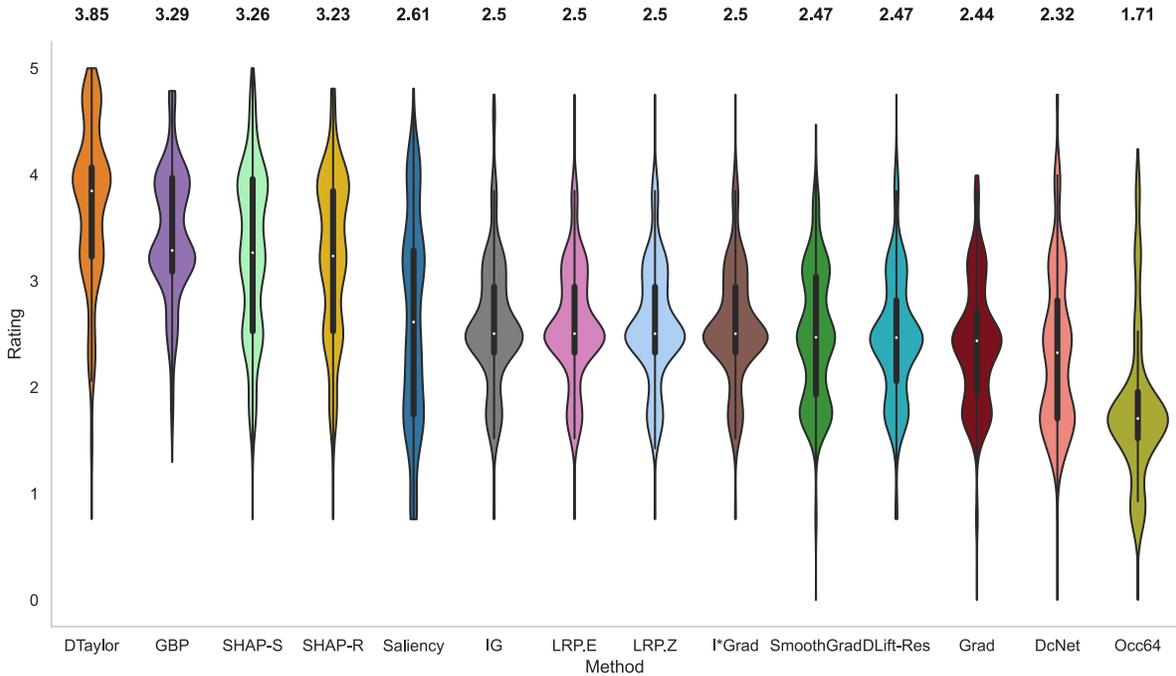

Figure 3: *Violin plots of normalized ratings of all methods. The breadth of the plot shows the probability density of the data and the median value is reported on top of the plots. Deep Taylor was rated the highest overall followed by GBP and SHAP.*

Table 2: *Median ratings (with range) for each disease for all attribution methods*

| Method | Median rating | | | |
|---|---|---|---|---|
| | **CNV** | **DME** | **Drusen** | **All** |
| DcNet | 2.17 (1.71-2.61) | 2.47 (1.74-3.09) | 2.32 (1.71-2.61) | 2.32 (1.71-2.82) |
| DTaylor | **3.80 (3.22-4.05)** | **3.48(3.09-3.99)** | **3.99 (3.58-4.56)** | **3.85 (3.23-4.07)** |
| DLift-Res | 2.44 (1.85-2.72) | 2.44 (1.96-2.53) | 2.53 (2.32-3.09) | 2.47 (2.06-2.82) |
| Grad | 2.32 (1.77-2.53) | 2.47 (2.19-2.95) | 2.44 (2.03-2.61) | 2.44 (1.96-2.72) |
| GBP | 3.23 (3.09-3.80) | 3.26 (3.07-3.80) | 3.71 (3.22-3.99) | 3.29 (3.09-3.97) |
| I*Grad | 2.50 (2.32-2.95) | 2.47 (2.28-2.82) | 2.53(2.44-3.04) | 2.50 (2.32-2.95) |
| IG | 2.50 (2.32-2.95) | 2.47 (2.19-2.82) | 2.57 (2.44-3.20) | 2.50 (2.32-2.95) |
| LRP.E | 2.50 (2.32-2.95) | 2.50 (2.32-2.95) | 2.53 (2.41-3.04) | 2.50 (2.32-2.95) |
| LRP.Z | 2.50 (2.32-2.95) | 2.50 (2.32-2.95) | 2.53 (2.41-3.04) | 2.50 (2.32-2.95) |
| Occ64 | 1.71 (1.55-1.96) | 1.71 (1.42-1.85) | 1.71 (1.42-1.96) | 1.71 (1.52-1.96) |
| Saliency | 2.47 (1.74-3.29) | 2.72 (1.74-3.29) | 2.61 (1.74-3.29) | 2.61 (1.74-3.29) |
| SHAP-R | 3.23 (2.53-3.85) | 3.23 (2.53-3.85) | 3.58 (2.89-3.96) | 3.23 (2.53-3.85) |
| SHAP-S | 3.23 (2.53-3.85) | 3.23 (2.53-3.85) | 3.53 (2.61-3.96) | 3.26 (2.53-3.96) |
| SmoothGrad | 2.45 (1.85-2.95) | 2.47 (1.96-3.09) | 2.47 (1.85-3.04) | 2.47 (1.93-3.04) |

## Comparison between raters

The Spearman's rank correlation was used to compare the ratings of the clinicians with each other. This test is a non-parametric measure that assesses the relationship between two variables, in this case the ratings of images by two different clinicians. A correlation of +1 indicates a perfect positive correlation, 0 indicates no correlation, and -1 indicates perfect negative correlation. The correlations between the ratings of all 14 clinicians for the 60 images and 13 methods are shown in figure 4. P1 to P10 are ophthalmologists while P11 to P14 are optometrists.

Most of the values are around 0.5 indicating an overall moderate agreement between clinicians. The highest correlation was of 0.76 between P10 and P13 while two cases of slight negative correlation were between P1 and P11 and P2 and P11. P10 and P13 are both optometrists whereas P1 and P2 are ophthalmologists and P11 is an optometrist who has relatively less experience with OCT and had lower correlation with all clinicians. This indicates that the background and training (i.e., prior experience) of clinicians affected their ratings of the system.

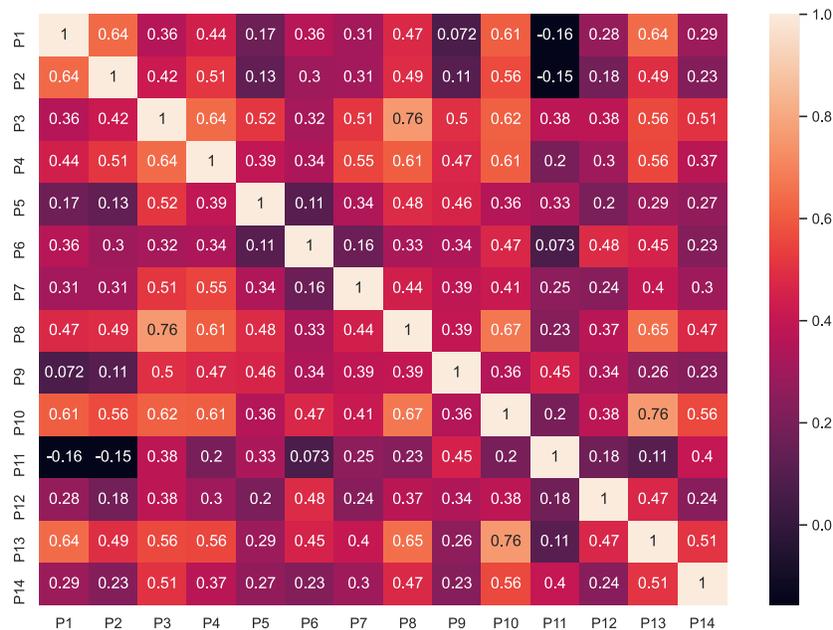

Figure 4: Spearman's correlation for clinician's ratings

## Qualitative observations

In this section, the qualitative feedback given by the clinicians regarding the performance of the system, potential use cases and other suggestions are summarized. A survey was collected from the clinicians to seek their opinion post study. It is notable that 79% (11/14) clinicians who participated in the study would prefer to have an explainable system assisting them in practice, reaffirming the need for such system to the clinical community. One of the ophthalmologists gave their feedback on the system as – "It is a definite boon to the armamentarium as far as screening and diagnosis is concerned on a mass scale or in a telemedicine facility".

The clinicians noted an overall better coverage of the pathology by Deep Taylor as the reason for higher ratings, however it and other methods except SHAP were found to be mainly detecting the boundaries. SHAP was observed to be identifying regions inside the edema also, though the partial coverage of the region lower score. The noise, (represented in blue) especially in case of LRP was found to be a distraction by some clinicians and can be removed for actual implementation.

Most of the clinicians identified telemedicine and tertiary care centres as potential sites which can utilize this system. It was suggested that it can be used for screening in places with large number of patients without sufficient number of specialists. It could help clinicians by categorizing the scans with suspect conditions and thus allows clinicians to focus their attention on examining the areas of the images highlighted by algorithm. This can improve efficiency and help in saving time, resulting in more efficient patient care. Another application could be archival and data management where the heatmaps could be used for separating images faster.

## DISCUSSION

Along with a comparison of various available attribution methods to explain deep learning models, this study validated their results through ratings from a large panel of clinicians. Most of them were not involved in the design process at this stage, were in general positive about the utility of the system and were and receptive to using this methodology.

A method based on Taylor series expansion, known as Deep Taylor, received the highest ratings showing that methods with stronger or better theoretical backgrounds and high performance on standard datasets may not be the optimal methods in a practical medical imaging situation. It must be noted that the original goal of these attribution methods was to explain the model's decision-making process by generating a true representation of the features used by a model to perform a given task. Hence, the heatmaps generated are affected by both by the model and the attribution method used. It must be noted that a significant issue with GBP, the second highest rated method is that it acts as an edge detector and not actually revealing the model's decision-making process [27,28]..

The dataset used here labeled only primary diagnosis, however, the clinicians were able to identify secondary diagnosis for some images from their evaluation. Also, due to the nature of the dataset the study is limited to a single orientation of the OCT scan which might differ between the images. All the clinicians in this study preferred having fundus images in addition to OCT, hence, a system that uses fundus, OCT, and patient data similar to [29] could be useful in practice. Another application of explainability system could be as a self-learning tool. The system can be further developed to encompass other diseases and finetuned for the specific imaging modality, taking into account variables such as noise, illumination, field position, etc. Currently, OCT systems are not used in eye camps for example since the devices are not only expensive but also are bulky and cannot be used for screening. Given recent advances in low-cost portable OCT devices [30], it is possible to integrate an explainable diagnosis system on a laptop or mobile device for teleophthalmology purposes and it would be invaluable to the clinical community.

## CONCLUSION

In this study multiple attribution methods for explaining OCT diagnosis in terms of heatmaps were compared by a panel of clinicians. This is to the best of our knowledge the first study to look at qualitative comparison of various explainable AI methods performed by a large panel of clinicians. A method based on Taylor series expansion, known as Deep Taylor, received the highest ratings outperforming the methods with stronger theoretical background and better results on standard datasets. Positive feedback about the use of such system was received from the panel of retina specialists. Future enhancements of the system could make it a trustable diagnostic assistant helping resolve the lack of access to ophthalmic healthcare.

## ACKNOWLEDGEMENTS

This work is supported by an NSERC Discovery Grant and NVIDIA Titan V GPU Grant to V.L. This research was enabled in part by Compute Canada (www.computecanada.ca). We acknowledge the team of vitreoretinal fellows from Sankara Nethralaya, Chennai for participating in the study. (Dr Janani Sreenivasan, Dr Rekha Priya Kalluri Bharat, Dr Shelke Komal Vishwambhar, Dr Kaushal Mukesh Sanghvi, Dr Anjalika Shekhar, Dr Akkarapaka Joash Rijay, Dr Subham Sinha Roy, Dr Isha Agarwal, Dr Krishna Kanta Roy, and Dr Puja Maitra).